\documentclass[pra,aps,twocolumn,showpacs,superscriptaddress]{revtex4}
\usepackage{graphicx}
\usepackage{amsmath,amsfonts}

\begin{document}


\title{Doorway states and the Bose-Hubbard model}
\author{A. N. Salgueiro}
\email{andrea.salgueiro@web.de} \affiliation{Instituto de
F\'{\i}sica, Universidade de S\~ao Paulo, CP 66318 CEP 05389-970,
S\~ao Paulo, Brazil}
\author{Chi-Yong Lin}
\affiliation{Department of Physics, National Dong Hwa University
Hua-Lien, Taiwan, R.O.C.}
\author{A.F.R. de Toledo Piza}
\affiliation{Instituto de F\'{\i}sica, Universidade de S\~ao Paulo,
CP 66318 CEP 05389-970, S\~ao Paulo, Brazil}
\author{M. Weidem\"uller}
\affiliation{Physikalisches Institut, Albert-Ludwigs-Universit\"at
Freiburg, 79104 Freiburg, Germany}





\begin{abstract}
We use the concept of doorway states to solve the finite
Bose-Hubbard model. This method allows for the inclusion of
many-body correlations in a dynamically motivated hierarchical
way, yielding useful approximations even within subspaces of the
full Hilbert space of greatly reduced dimensionality.
Moreover we apply the doorway method to solve the problem of $N$
bosons in a lattice, where the chemical potential, the on-site
fluctuations, the visibility of the interference pattern arising
from atoms in a one-dimensional periodic lattice and the width of
the interference peak are calculated. Excellent qualitative
agreement with exact numerical calculations as well as recent
experimental observations is obtained.
\end{abstract}

\maketitle

\section{Introduction}

The physics of strongly interacting quantum systems has been
subject of intense investigations since the early days of quantum
mechanics. With the recent developments in the field of ultracold
quantum gases, systems of strongly correlated bosonic and
fermionic atoms with tunable interactions have become amendable to
experimental studies
\cite{ketterle,grimm,immanuel,kasevich,esslinger,ober}. Complete
theoretical understanding of these coupled many-body systems is
difficult, and only rare examples exist, for which analytical
solutions for the ground state could be explicitly been given
\cite{many}. Of particular importance in this context is the
Bose-Hubbard model, which can describe bosons or fermions in an
optical lattice or in a double well potential. This model serves
as a prototype system exhibiting a quantum phase transition
\cite{fisher}.
Triggered by the recent observation of this phase transition from
a superfluid to a Mott-insulating phase with bosonic atoms in one-
(1D) and three-dimensional (3D) optical lattices
\cite{kasevich,immanuel, tilman}, the ground state and correlation
properties of the Bose-Hubbard model have been extensively
investigated by a large number of groups \cite{roth,burnett,mf}.
Complete theoretical understanding of these coupled many-body
systems is difficult, and only rare examples exist, for which
analytical solutions for the ground state could be explicitly been
given \cite{many}.
The description generally has to rely upon approximation of
numerical or analytical methods such as the
density-group-renormalization \cite{DMRG}, quantum Monte Carlo
approaches \cite{monte}, perturbative calculations \cite{monien},
and mean-field treatments ,e.g., Gutzwiller ansatz
\cite{gutzwiller1,gutzwiller2,bruder}. In this context, it is
necessary to introduce different methods which give the energy
spectrum of a many-body hamiltonian and its respective eigenstates
or at least the ground-state.

In this paper, we use the concept of doorway states to study
bosons in a double well and bosons in an optical lattice. Both
systems can be described by the Bose-Hubbard model. The method is
based on the construction of doorway projectors as initially
introduced by Feshbach in the context of scattering theory in
Nuclear Physics to explain the appearance of intermediate
structures in the cross section \cite{feshbach}. The main idea
underlying this approach is a dynamically motivated decomposition
of the full Hilbert space of the hamiltonian into a sequence of
mutually orthogonal subspaces. The first subspace in the hierarchy
is chosen on the basis of its particular relevance to the problem
at hand and making use of the symmetries of the hamiltonian. In
the original context of nuclear reactions this subspace spans the
open channels. In the context of the Bose-Hubbard model, it spans
the analytical lowest energy solutions of the pure hopping or the
pure superfluid limiting cases, when one is interested in
approximating the system ground-state.
Subsequent subspaces are chosen as spanned by new state vectors
generated from the action of the hamiltonian on each partially
defined subspace. Formally, this is implemented by establishing a
sequence of mutually orthogonal projectors $\mathbb{P},
\mathbb{D}_1, \mathbb{D}_2, \cdots$ such that $\mathbb{P}+\sum_i
\mathbb{D}_i$ is the identity operator and the hamiltonian has
non-vanishing matrix elements only between states in consecutive
members of the hierarchy. This hierarchy allows for dynamically
controlled truncation schemes which may significantly reduce the
computational effort. The advantage of the doorway approach is
that it can be used as a numerical or analytical tool. The doorway
method is also know as projection formalism and continued-fraction
representation \cite{ziegler}. Moreover, this approach can be seen
as a generalization of the several other schemes developed in
different areas of many-body physics, such as the Lanczos method
\cite{lanczos} of Solid State physics, the resolvent method
\cite{resolvent} of Quantum Optics, and the Feshbach-Fano
partitioning method \cite{fano-feshbach} of Quantum Chemistry.
Note, in particular, that in each of these methods the subspaces
are one-dimensional.

Usually it is much more difficult to obtain the eigenstates of a
given hamiltonian rather than the eigenvalues of the energy. In
this context, the way the doorway method is constructed creates a
reduced subspace where the Bose-Hubbard hamiltonian can be
projected. The projection of the hamiltonian in this reduced
subspace allows for the knowledge of part of the energy spectrum
and its correspondent eigenstates. We are mainly interested in the
ground-state of this system which features both the superfluid and
Mott-insulator regimes. In this context, the energy spectrum with
the same symmetry of the ground-state and the ground-state
wavefunction of the Bose-Hubbard model are calculated.
Furthermore, the correlation properties of a low density bosonic
system in an optical lattice are calculated over the full range of
coupling parameters and found to qualitatively describe the
measurements of \cite{tilman,bloch1}, even though these
experiments are inhomogeneous systems due to the presence of an
external trap.

The paper is organized as follows. In section \ref{secum} a more
formal presentation of the doorway approach is given. In section
\ref{secdois} the doorway approach is applied to solve $N$ bosons
distributed in $M$-sites, described by the Bose-Hubbard model. In
subsection \ref{secdoisa} we start with $M=2$ sites and the
doorway method is used to obtain the energy spectrum and the
ground-state. In subsection \ref{secdoisb} the approach is applied
to the more general case ($N$-boson distributed among $M$-sites),
where experimental observables such as chemical potential, on-site
variance and visibility are calculated.

\section{Description of the doorway method}
\label{secum}

The doorway method consists in splitting the full Hilbert space
of the many-body hamiltonian $H$ into different
subspaces which are accessed by the action of projectors on $H$. 
Let $P$ be the projector onto the subspace $\mathbb{P}$ generated by
a set of state vectors which are relevant for the lowest order
approximation to the system at hand. Specific choices in connection
with Bose-Hubbard systems are discussed in the subsections below.
$\mathbb{P}$ is in general multi-dimensional, so that
$P=\sum_{i=1}^k |p_i\rangle \langle p_i|$ in terms of a orthonormal
set of state vectors $\{|p_i\rangle\}_{i=1,\cdots k}$.
From $P$, it is possible to construct the basis which generates
the first doorway subspace by the action of $|d_i^{(1)}\rangle
\equiv(1-P)H|p_i\rangle$ which have to be normalized. The second
doorway state is given by
$|d_i^{(2)}\rangle\equiv(1-P-D_{1})H|\tilde{d}_i^{(1)}\rangle$,
with $|\tilde{d}_i^{(1)}\rangle$ as the normalized first doorway
state. The second doorway $|d_i^{(2)}\rangle$ has also to be
normalized. The orthogonalization of the doorway states is done at
each step. The procedure terminates when $(1-P-D_{1}-D_{2}-
\cdots-D_{k})H|d_i^{(n-1)}\rangle=0$ with $D_{j} =\sum_i^k
|\tilde{d}_i^{(j)}\rangle\langle \tilde{d}_i^{(j)}|$
($j=1,2,\dots,n$) denoting the projectors onto the doorway spaces.
The space $\mathbb{P}\oplus\mathbb{D}_{1}\oplus \cdots \oplus
\mathbb{D}_{n}$ then represents an irreducible subspace for the
Hamiltonian containing the initial subspace $\mathbb{P}$. The
crucial point is the choice of the most adequate starting state
$|p_i\rangle$ to generate the initial subspace to obtain the
solution in the most efficient way. The many-body Schr\"odinger
equation is solved in this subspace

\begin{eqnarray}
\label{dois}
(E-H_{PP}-H_{PD_{1}}\frac{1}{E-H_{D_{1}D_{1}}-M_{1,2}}H_{D_{1}P})
P|\psi\rangle=0, \\
M_{i,i+1}\equiv
H_{D_{i}D_{i+1}}{(E-H_{D_{i+1}D_{i+1}}-M_{i+1,i+2})}
^{-1}H_{D_{i}D_{i+1}} \nonumber
\end{eqnarray}

\noindent where $H_{PP}=PHP$, $H_{D_i D_j}=D_{i}HD_{j}$,
$H_{D_{i}P}=D_{i}HP$ with $i,j = 1,2\dots n$. Using the second
equation recursively, one obtains a nonlinear problem within the
$P$-subspace which determines the eigenvalues of the hamiltonian
in the full Hilbert space. All orders of interactions described by
the Hamiltonian are taken into account inside this subspace.
Therefore, the expansion is not perturbative in the couplings, but
it represents the full solution in a subspace of lower (reduced)
dimension. Once the solution is truncated, the diagonalization of
the hamiltonian happens within the used subspace (e.g.
$\mathbb{P}\oplus\mathbb{D}_{1}\oplus \cdots \oplus
\mathbb{D}_{last{\quad}used}$).

Information about the eigenstates of the system can be obtained by
projecting the original hamiltonian in the reduced subspace
created by the doorway states, which can be either complete or
truncated, depending on the physical system. In this case the
quantities $\tilde{H}_{PP}=\langle P|H_{PP}|P\rangle$,
$\tilde{H}_{D_i D_j}=\langle D_{i}|H_{D_i D_j}|D_{j}\rangle$,
$\tilde{H}_{D_{i}P}=\langle D_{i}|H_{D_{i}P}|P\rangle$ are the
matrix elements of the reduced hamiltonian. In spite of the
original doorway procedure giving no information about the
eigenstates of the system, one can use the reduced subspace which
is created by the doorway states to project the hamiltonian in
this reduced subspace, and then diagonalize it to obtain
information about the eigenstates of the system.

\section{Doorway method applied to the Bose-Hubbard model}
\label{secdois}

We apply the doorway method to
a  system of $N$ bosons in a 1D optical lattice of $M$ sites in the
homogeneous case. The system is described by the Bose-Hubbard model
\begin{equation}
\label{um} H=-J\sum_{j=1}^M(a_j^{\dagger} a_{j+1}+h.c.)+\frac{U}{2}
\sum_{j}n_{j}(n_{j}-1)
\end{equation}

\noindent which can be seen as a many-mode approximation for an
ultracold gas of bosonic atoms occupying the lowest band in a
periodic array of potential wells \cite{fisher,gutzwiller2}. The
operators $a_j^{\dagger}$ and $a_{j}$ are the creation and
annihilation operators, respectively, of a boson in the $j$-th site,
and $n_{j}$ is the number of atoms at the lattice site $j$
($j=1\ldots M$). The strength of the on site repulsion of two atoms
on the lattice site $i$ is characterized by $U$. The parameter $J$
is the hopping matrix element between adjacent sites which
characterizes the strength the tunneling term. A periodic boundary
condition is included allowing for tunneling between the first and
the last site of the lattice (site $M+1 \equiv$ site 1). In the case
of $M=2$ sites, the periodic boundary condition is already
intrinsic.

In this paper, we consider the Bose-Hubbard hamiltonian with
periodic boundary conditions. 
Moreover, the Bose-Hubbard hamiltonian
with periodic boundary conditions conserves two important
quantities: the total number of particles and the total
quasi-momentum. As a consequence, its Hilbert space can be
partitioned into M decoupled subspaces corresponding to the values
of quasi-momentum (q=0,...,M-1). Furthermore, the subspace with q=0
and, for even M, also the subspace with quasi-momentum q=M/2, can be
further partitioned into symmetric and antisymmetric subspaces
("odd-even" symmetry) \cite{buchleitner}. The ground state of the
system lies in the symmetric  $q=0$ subspace. Thus, the initial P
subspace is chosen within this subspace. The full set of doorway
subspaces constructed from the initial $\mathbb{P}$ subspace will
finally span the symmetric part of the subspace with zero
quasi-momentum, providing the full solution for the ground state. In
order to define the $\mathbb{P}$ subspace, one can use the fact that
the ground state of the system has two different limiting cases: the
superfluid phase for $J \gg U$ where the particles are delocalized
over the whole lattice, and the Mott-insulating phase for $J\ll U$
where the particles are localized with an equal number of atoms per
site $n_{0}=N/M$.

\subsection{Two-site Bose-Hubbard model}
\label{secdoisa}

To illustrate how the method works, we first apply the doorway
method to the two-site Bose-Hubbard model, which corresponds to
$N$ bosons in a double-well. First, the doorway method is used to
construct the reduced subspace the two-site Bose-Hubbard
hamiltonian is projected. The projected hamiltonian can be
diagonalized. Analytical expressions for the matrix elements of
the reduced hamiltonian $\tilde{H}_{pp}$, $\tilde{H}_{pd_i}$,
$\tilde{H}_{d_{i}d_{i}}$ and $\tilde{H}_{d_{i}d_{i+1}}$ are found.
We start the doorway construction for the repulsive case using
different $|p_i\rangle$ states, since there is a crucial
difference between the ground state for $N$-even and $N$-odd.  For
$J=0$, the $N$-even ground state is non-degenerate,
$|p\rangle=|(N/2)(N/2)\rangle$, where the number of particles per
site is an integer (Mott-insulator) while the $N$-odd ground state
is degenerate $|p_{\pm}\rangle=1/\sqrt{2}[|(N+1)/2,
(N-1)/2\rangle\pm|(N-1)/2, (N+1)/2\rangle]$. In the first case $P$
is one-dimensional and in the second case $P$ is bi-dimensional.
Now that $P$ projectors are defined, the doorway projectors can be
built. For $N$-even the total doorway projector is given by
$D=\sum_{i=1}^{l} D_{i}$, where $l=N/2$ and generator state of the
doorway subspace is $|d_{i}\rangle=(1/\sqrt{2})[|(N/2+i)
(N/2-i)\rangle+|(N/2-i) (N/2+i)\rangle]$, where $i$ is the number
of particles one can remove or add to a particular site ($1\leq
i\leq N/2$). For $N$-odd the doorway projector is given by
$D=\sum_{i=1}^{l} (D_{i+}+D_{i-})$, where $l=(N-1)/2$ and the two
unconnected type of generator states of the doorway subspace are
related to the symmetric and antisymmetric subspace, namely
$|d_{i\pm}\rangle=(1/\sqrt{2})[|((N+1)/2+i),
((N-1)/2-i)\rangle\pm|((N-1)/2-i),((N-1)/2+i)\rangle]$, where $i$
is the number of particles one can remove or add to a particular
site ($1\leq i\leq (N-1)/2$). 
Now that we know all doorway projectors and $P$, the Bose-Hubbard
hamiltonian can be projected in the subspace generated by them and
the matrix elements of the projected hamiltonian $\tilde{H}_{pp}$,
$\tilde{H}_{pD_{1}}$, $\tilde{H}_{D_{i-1}D_{i}}$ and
$\tilde{H}_{D_{i}D_{i}}$ where $1\leq i\leq n_{0}=l$, with $l=N/2$
for $N$
even and $l=(N-1)/2$ for $N$ odd, can be calculated. 
For $N$-even,

\begin{eqnarray}
&&\tilde{H}_{pp}=\frac{U}{2}N(\frac{N}{2}-1)|p\rangle\langle
p|\\\nonumber
&&\tilde{H}_{Pd_{1}}=-J\sqrt{N(N/2+1)}|p\rangle\langle
d_{1}|\\\nonumber &&
\tilde{H}_{d_{i}d_{i}}=[\frac{U}{2}N(\frac{N}{2}-1)+Ui^2]|d_{i}\rangle\langle
d_{i}|\\\nonumber &&
\tilde{H}_{d_{i-1}d_{i}}=-2J\sqrt{(\frac{N}{2}+i)(\frac{N}{2}-i+1)}|d_{i-1}\rangle\langle
d_{i}|.
\end{eqnarray}

\noindent while for $N$-odd

\begin{eqnarray}
&&\tilde{H}_{pp}=[\frac{U}{4}{(N-1)}^2-\frac{J(N+1)}{2}]|p_{+}\rangle\langle
p_{+}| \nonumber \\
&&+[\frac{U}{4}{(N-1)}^2+\frac{J(N+1)}{2}]|p_{-}\rangle\langle
p_{-}|\nonumber\\ &&\tilde{H}_{d_{i}d_{i}}=[\frac{U}{4}{(N-1)}^2+U
i(i+1)][|d_{i+}\rangle\langle d_{i+}|+|d_{i-}\rangle\langle
d_{i-}|] \nonumber\\
&&\tilde{H}_{d_{i-1}d_{i}}=-(J/2)\sqrt{({(N+1)}^2-4i^2)}[|d_{(i-1)+}\rangle\langle
d_{i+}|\nonumber\\ && +|d_{(i-1)-}\rangle\langle d_{i-}|].
\end{eqnarray}

\noindent Consequently the diagonalization of the reduced
hamiltonian can be performed. 
The doorway solution of the energy spectrum of the system is, of
course, in perfect agreement with the exact diagonalization of the
hamiltonian. The doorway solution of the energy spectrum, which
corresponds to the exact one, can be fully generated by
considering $N/2+1$ ($N$ even or odd) doorways for the symmetric
and $N/2$ ($N$ even) or $N/2+1$ (for $N$ odd) doorways for the
antisymmetric situation. It is well-know that the energy spectrum
of this system is composed of regions with and without
quasi-degenerate doublets \cite{wall} which is confirmed by the
doorway solution. The degeneracy of the doublets is broken as the
tunneling parameter increases. The same analysis can be extended
to the case of attractive on-site interaction.


In the case considered here, the doorway method is optimized to
give information about the eigenstates of the system, but the
attention will be
focus on the ground-state/first excited state and the last two excited states of the system. 
The general form of the ground state of this system with $N$-even
particles as a function of the tunneling parameter is
$|\psi_0\rangle=\alpha_p|p\rangle+\sum_{i=1}^{N/2}\alpha_{d_i}|d_{i}\rangle$.
For this particular case, the ground-state is the analogous to the
Mott state for the two-site Bose-Hubbard model. This ground-state
is an unique state. However, the ground state of the system with
$N$ odd is a degenerate doublet for $J/U=0$. This degeneracy of
the ground state is lifted as the tunneling parameter increases.
Thus, one of the states of the quasi-degenerate doublet becomes
the ground-state while the other state becomes the first excited
state.
The ground-state and the first excited state are given
respectively by
$|\psi_0\rangle=\alpha_{P_{+}}|P_{+}\rangle+\sum_{i=1}^{(N-1)/2}\alpha_{d_{i+}}|d_{i+}\rangle$
and
$|\psi_1\rangle=\alpha_{P_{-}}|P_{-}\rangle+\sum_{i=1}^{(N-1)/2}\alpha_{d_{i-}}|d_{i-}\rangle$,
as a function of the tunneling parameter. We compare the form of
the wave-functions discussed above with the semi-classical
solution. The semi-classical solution does not distinguish if $N$
is even or odd as the quantum one does. The semi-classical
solution is always localized (see ref.\cite{piza}). It also does
not make any difference if the on-site interaction is repulsive or
attractive.

The last two excited states are a quasi-degenerate doublet, where
one of the states of the doublet. For $J/U=0$, the last two
excited states are given by
$|\psi_{e_{\pm}}\rangle=1/\sqrt{2}[|N0\rangle\pm|0N\rangle]\equiv|d_{last}\rangle$
and for $J/U\neq0$ new components are created in the states
$|\psi_{e_{\pm}}\rangle=\alpha'_{P_{\pm}}|P_{\pm}\rangle+\sum_{i=1}^{(N-1)/2}\alpha'_{d_{i\pm}}|d_{i\pm}\rangle$.
These excited states represent even-odd macroscopic superpositions
(even and odd Schr\"odinger cat states). Nowadays, enormous effort
has been put to create such states, since they are important to
quantum information, spectroscopy and so on. Those states have
been theoretically found before as the symmetry-preserving-class
of solutions of the GP equation which preserves the symmetry
\cite{reinhardt3}.
It has been observed these even-odd macroscopic superpositions do
not lose their strength until $J/U=1$, and after that point, they
start to disappear very quickly. To understand why they are more
robust than any other states of spectrum, one has to look how the
energy spectrum changes as a function of the tunneling parameter.
These states are the last to be affected by the increase of
tunneling parameter. It is the last degeneracy to be broken.
 even with a considerable vale of the
tunneling parameter $J/U$.

\subsection{Bose-Hubbard model for a lattice} \label{secdoisb}

We extend the previous analysis to a more general case, the
Bose-Hubbard model for a lattice ($M$-sites). Moreover,
observables of one-body and two-body, such as chemical potential,
variance and visibility are calculated with the help of the
doorway method. For the commensurable situation ($N=n_0 M$) we
choose $|p\rangle=|n_{0}\rangle^{\otimes M}$ corresponding to the
ground state of the system in the pure Mott case ($U \gg J$),
which generates a one-dimensional subspace $\mathbb{P}$. If an
extra particle is added to or removed from the system ($N\pm
1=n_{0}M\pm 1$) (non-commensurate situation), the initial
multi-dimensional subspace $\mathbb{P}$ is generated and has the
form $\mathbb{P}=\sum_{i=1}^M|p_i\rangle\langle p_i|$, where
$|p_i\rangle =|(n_{0}\pm1)_in_{0}^{\otimes M}\rangle$. Starting
from these initial subspaces the correspondent doorway subspaces
are constructed. For non-commensurate situation, however there is
a simpler way to obtain only the ground-state, which is to start
instead with the symmetric superposition  $|p\rangle =\frac{1}{
\sqrt{M}}\sum_{j=1}^{M}|(n_{0}\pm1)_jn_{0}^{\otimes M}\rangle$.
Furthermore, the doorway method gives the flexibility to choose
another states as the starting point $|p\rangle$. Depending on the
parameters of interest, one may also start from the pure
superfluid state ($J \gg U$) as the initial state $|p\rangle$ in
the construction of the doorways, following the same procedure
discussed in the section before. We will use this flexibility to
study the on-site variance later. After finishing to generate the
doorway subspaces, the part of the energy spectrum and the
ground-state are obtained by solved by projecting the $M$-site
Bose-Hubbard hamiltonian in the reduced subspace created by $P$
and doorway projectors. Unfortunately for this general case, it is
not easy to obtain analytical expressions for the matrix elements
of the reduced hamiltonian as in the previous case ($M=2$). The
exact ground-state of the system can be described in general as
$|\psi_0\rangle=\sum_{i=0}^{n}\alpha_i|d_i\rangle$, where
$|d_0\rangle=|p\rangle$, where $n$ is the number of last generated
doorway. One can check that the superfluid ground state
$|\psi_{\mathrm
SF}\rangle\sim{(\sum_{i=1}^{M}a_j^{\dagger})}^N|0\rangle$ is
asymptotically reached for $J/U\rightarrow\infty$.

\begin{figure}
\centering
\includegraphics[width=0.9\columnwidth]{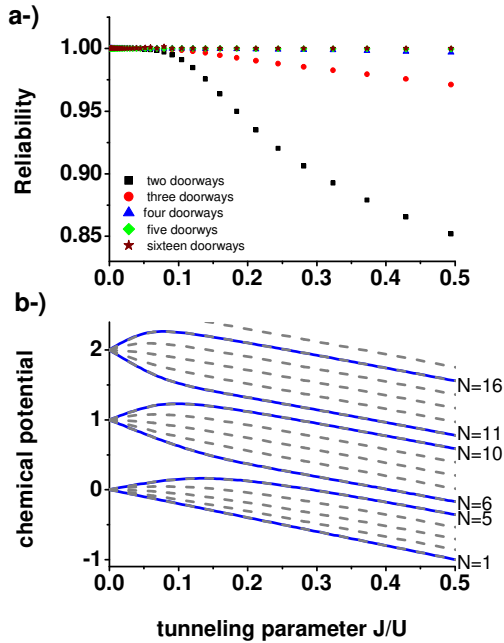}
\caption{(a) Reliability as a function of the tunneling parameter
for N=M=5 and the maximum number of doorways $n=2,5,16$.
(b)Chemical potential $\mu_N/U$ as a function of the dimensionless
tunneling parameter $J/U$ for $M=5$ lattice sites and $N$
particles. The solid lines are the result of the doorway method
using seven doorways. The dashed curves represent an exact
numerical solution of the Bose-Hubbard model.} \label{one}
\end{figure}

The doorway method genuinely reduces the numerical effort to the
diagonalization of the very small, but relevant part of the total
Hilbert space. For small number of sites, it is possible to obtain
numerically the full solution of the energy spectrum and ground
state. However, for larger number of sites, it is not so easy to
generate all doorways one needs to obtain the full solution
because of numerical reasons, such as lack of available memory. In
this context it is important to discuss the convergency of the
method. A good measure for the convergence of the method is given
by the "reliability" defined as
$\mathcal{R}={\sum_{i=0}^{n'}|\langle\psi_0|d_i\rangle|}^2$, where
$|\psi_0\rangle$ being the exact ground state of the system
obtained from the exact numerical diagonalization of the
Bose-Hubbard Hamiltonian and $n'$ being the maximum number of
doorways to generate the truncated solution. The meaning of the
reliability can be understood in terms of the portion of the full
symmetric subspace which has been covered by the truncated
solution. The convergency of the method scales linearly with the
number of particles for the commensurate and incommensurate cases.
We have tested this convergency for cases we are able to obtain
the exact solution. Fig.\ref{one}(a) illustrates this convergency
for $M=N=5$, considering the truncated solution for $n=2,3,2,5,16$
doorways. The convergency is then achieved after the fifth doorway
$|d_5\rangle$ is created, showing the subspace spanned by the
doorways already include the most relevant portion of the full
symmetric subspace. For this particular case, the Hilbert space is
of dimension $\binom{N+M-1}{N}=126$ whereas the symmetric zero
quasi-momentum subspace has only dimension $16$.
The convergency of the method is surprisingly fast, since the
recursive construction of the doorway states ``automatically''
creates the most efficient path into the irreducible subspace
depending on the specific values of the parameters $U$ and $J$.



\paragraph*{Chemical potential:}
We derive the ground state energy and the chemical potential
($\mu_N/U=[E_{N+1}-E_N]/U$) as a function of the dimensionless
tunneling parameter $J/U$. Fig.~\ref{one}(b) shows the chemical
potential as a function of $J/U$ comparing the exact solution of the
Bose-Hubbard hamiltonian (dashed lines) with the result of the
doorway method (solid lines) for the case of $M=5$ sites. The
distance between the apparent parallel lines of Fig.~\ref{one}(b) is
almost $1/M$, which vanishes in the thermodynamic limit
($N,M\rightarrow\infty$ and $N/M$ finite), characterizing the phase
transition point where the two limiting curves $\mu_{n_0 M}$ and
$\mu_{n_0 M+1}$ are the same. For convenience, the doorway method
has only been applied to the limiting cases $N=n_0 M$ and $N=n_0 M
\pm1$ where the determination of the corresponding initial states is
described above.
The curves inside the lobes are only shown for the exact calculation
and they represent the evolution of the $M$-degenerate states of
$J=0$ as a function of the hopping parameter.

\begin{figure}
\centering
\includegraphics[width=\columnwidth]{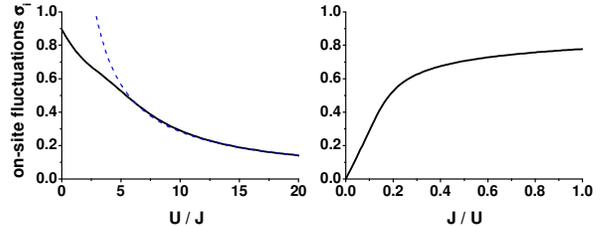}
\caption{On-site number fluctuations
calculated for $M=6$ sites and $N=6$ particles using eight
doorways to construct the ground state of the system. The initial
state in the recursion is either the pure superfluid state (left
graph) or the pure Mott-insulator state (right graph). The dashed
line gives the result of perturbation theory for $J/U\ll 1$.}
\label{two}
\end{figure}

\paragraph*{On-site number fluctuation:}
As a second application of the doorway method, we determine the
on-site number fluctuation of the ground state
$\sigma_i=\sqrt{\langle n_i^2\rangle-{\langle n_i \rangle}^2}$ as a
function of $U/J$ and as a function of $J/U$. The results are shown
in Fig.~\ref{two}. Depending on the regime, the most efficient
construction of the doorways starts from either a pure superfluid
state (left graph in Fig.~\ref{two}) or the Mott state (right graph
in Fig.~\ref{two}). For small $U/J$, the on-site fluctuations are
found to decrease linearly starting from the initial value given by
$\sqrt{\frac{N}{M}(1-\frac{1}{M})}$. At intermediate $U/J$, the
decrease curve changes slope in the region of the phase transition
and finally approaches the dependence
$2\frac{J}{U}\sqrt{n_0(n_0+1)}$, which is identical to the
perturbative result for the ground state wavefunction
$|\psi\rangle\simeq|n_{0}\rangle^{\otimes
M}+\frac{J}{U}\sum_{\langle i,j \rangle} a_i^\dagger a_j
|n_{0}\rangle^{\otimes M}$ with $\langle i,j \rangle$ denoting the
next neighbor sites. The results of the doorway method agree well
with an analysis by Roth and Burnett based on the calculation of the
superfluid fraction under twisted boundary
conditions~\cite{roth}. 
Starting from the pure Mott-insulator state, we derive the on-site
number fluctuations for $J/U <\sim 1$ (right graph in
Fig.~\ref{two}). In the pure Mott phase, no correlations among the
particles exits and the one-site number fluctuations vanish. As
$J/U$ increases, correlations are created resulting in a steady
increase of the fluctuations. After the phase transition region is
crossed, the fluctuations approach the asymptotic value for $U/J=0$.
It is instructive to compare the result with the commonly used
Gutzwiller ansatz ~\cite{gutzwiller2,bruder} which, by construction,
does not provide particle correlations and thus genuinely
underestimates on-site fluctuations at small $J/U$. The short-range
correlations of the particles have to be included by additional
pertubative terms \cite{mf,bruder}. Since the groundstate
wavefunction constructed by the doorway method contains all relevant
many-body correlations, the on-site fluctuations are rendered
correctly for small $J/U$ as well as for the region of the phase
transition at intermediate values of the tunneling parameter.

\paragraph*{Interference pattern and visibility:}

\begin{figure*}
\includegraphics[width=1.4\columnwidth]{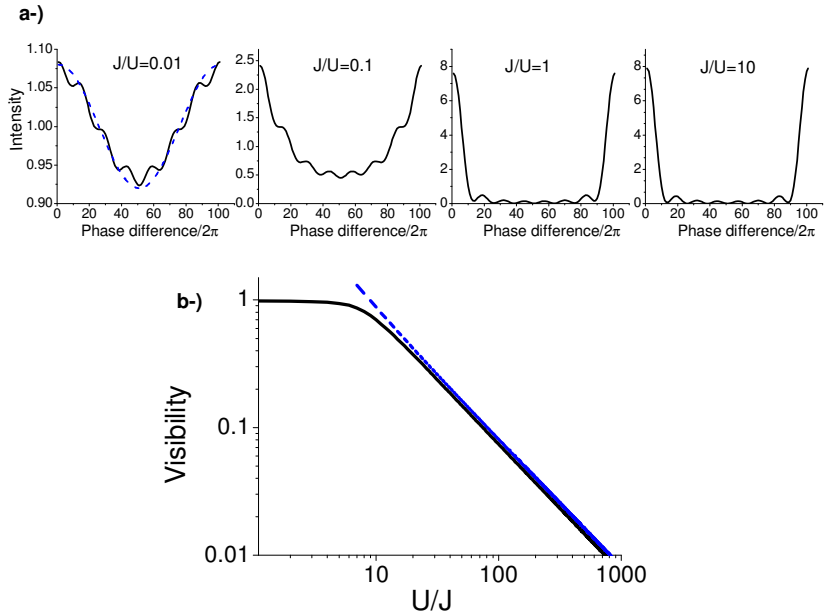}
\caption{(a) Interference pattern of a matter waves released from
a one-dimensional lattice for different values of the tunneling
parameter $J/U$ for eight particles distributed over eight lattice
sites. The intensity is plotted as a function of the accumulated
difference in phase after the expansion. (b) Visibility of the
interference pattern as a
function of the inverse tunneling parameter $U/J$. 
The dashed lines in (a) and (b) show the result of perturbation
theory for an infinite number of sites but finite average
population per site $n_0=1$.} \label{three}
\end{figure*}

To demonstrate the efficiency of the doorway method for the
determination of important experimental observables, the
interference of atoms released from an optical lattice is modeled
and the dependence of the visibility and the width of the first
resonance peak of the matter-wave interference pattern on the
parameter $J/U$ is derived. Comprehensive studies of these two
quantities have recently been performed by Gerbier \textit{et al.}
\cite{bloch1,bloch2} for 3D optical lattice and by St\"oferle
\textit{et al.} as well as M. K\"ohl \textit{et al.} for 1D optical
lattice \cite{tilman}. The observed interference pattern is directly
related to the Fourier transform of the single-particle density
matrix $\rho_{ij}=\langle a^{\dagger}_j a_i\rangle$. Therefore,
measurements of the visibility are particularly sensitive to
short-range coherence induced by the tunneling of the particles.
Following the derivation in Refs.~\cite{roth,kashurnikov}, the
intensity is calculated for a one-dimensional lattice as
$I=\frac{1}{M}\sum_{i,j=1}^M\exp\left(i k (r_i-r_j)\right)\langle
a_j^\dagger a_i\rangle$ where $k$ denotes the wave vector of the
expanding matter wave after release from the lattice point $r_i$ or
$r_j$, respectively. The result of the doorway method for $N=8$
particles distributed over $M=8$ sites, starting from the pure Mott
state and taking nine doorways into account, is shown in
Fig.~\ref{three}. Periodic boundary conditions assure that limited
size effects are minimized. Fig.~\ref{three}(a) depicts the
resulting intensity distribution for different tunneling parameters.
One clearly sees the development of sharp multiple-beam interference
maxima with increasing $J/U$, corresponding to a decrease of the
well depth in the experiments. For very small $J/U$, the remaining
weak modulation of the intensity distribution indicates small-scale
correlations even deep in the Mott-insulator regime, as have been
observed experimentally ~\cite{tilman,bloch1,bloch2}. The
intensity pattern for small $J/U$ is well fitted by the result of
perturbation theory for an infinite number of sites given by $I=1+8
\frac{J}{U} \cos (k d)$ (dashed line in Fig.~\ref{three}) where $d$
is the separation of two adjacent lattice sites.

In Fig.~\ref{three}(b) the visibility, defined as
$V=(I_{\mathrm{max}}-I_{\mathrm{min}})/(I_{\mathrm{max}}+I_{\mathrm{min}})$,
is plotted over a wide range of the parameter $U/J$. The power of
the doorway method becomes apparent by the fact that all features
of the visibility are correctly reproduced over the full range of
$U/J$ by using a ground state wavefunction created by the
superposition of few doorway states including the initial
Mott-insulating state. For large $U/J$ (Mott-insulator regime),
the visibility approaches the non-vanishing perturbative result
$V=4 (n_0+1)J/U$ \cite{bloch1,Yu,bloch2} reflecting the
persistence of short-range correlations. In the ballpark of
the phase transition 
the visibility changes its analytical dependence on $J/U$ and
finally reaches a value close to unity for small $U/J$.
This graph qualitatively describes 
the experimental findings of Gebier \textit{et al.}~\cite{bloch1},
even though the present model neglects the external trap and it is
not 1D. A detailed investigation of the visibility in a 3D-
inhomogeneous case  can be found in Ref. \cite{vis}.

\begin{figure}
\includegraphics[width=0.9\columnwidth]{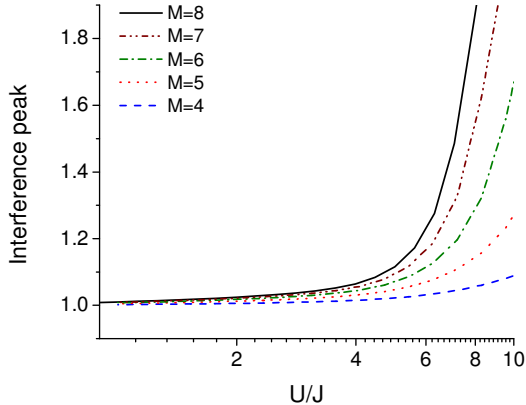}
\caption{The width of the first interference peak as
a function of $U/J$ for different number of sites $M=4,5,6,7,8$. 
} \label{four}
\end{figure}

Moreover, we can
directly calculate the width of the first interference peak for
different number of sites and compare with the experimental
measurements of Refs.~\cite{tilman} for 1D optical lattice. In
Fig.~\ref{four}, the width of the first interference peak is
calculated using the doorway method for different number of sites
($M=3,4,5,6,7,8$). We observe that as the number of sites increases,
the point the width curve starts to grown converges to a critical
value which characterize the phase transition region. Our
calculation describes qualitatively the results of refs.
\cite{tilman} until the phase transition point is reached. On the
other hand, after this point is crossed we see in our model the
finite size effects which does not show up in the experimental
results. Moreover, we do not observe the presence of kinks neither
in the visibility or in the width of the interference peak, since
the external trap is neglected in the present model.

\section{Conclusions}
\label{sectres}

In conclusion, we optimize the doorway states to solve the
Bose-Hubbard model for interacting bosonic particle in a periodic
lattice. The method is based on the successive construction of
doorway states which genuinely takes profit of symmetries included
in the hamiltonian. It naturally terminates when the full
irreducible subspace containing the starting state is spanned by
the doorway states. The intermediate doorway states depend on the
interaction and tunneling parameters of Bose-Hubbard hamiltonian
providing the closest approximation to the ground state.
Therefore, the convergence of the method, as can be quantified by
the ``reliability'' introduced in this paper, scales as the number
of particles.
All relevant correlations of the many-body system are included in
the ground state wavefunction derived in this way, as was shown by
calculating the energy spectrum, the on-site fluctuations and the
expectation value for single-particle correlations. We have limited
our discussion on a one-dimensional model to keep the calculational
effort on a low level. Conceptually, there are no limitations for
extending the method to 2D, 3D and to the inhomogeneous case. Other
important observables currently under experimental investigation,
such as two-particle correlations of Hanbury-Brown and Twiss
type~\cite{HBT}, can also be approached by the doorway method. Since
the doorway method allows direct access to the many-body
wavefunction of the system, it can be applied to other important
systems involving many-body correlations such as fermionic gases
coupled by Feshbach resonances or the fermionic variant of the
Bose-Hubbard model. In addition, dynamical properties of the
Bose-Hubbard model, such as many-body tunneling rates, may be
investigated.

\acknowledgments
We thank W. Zwerger, A. Buchleitner, A. Malvezzi,
T. St\"oferle and T. Esslinger for enlightening discussions. In
particular we thank T. St\"oferle and T. Esslinger to provide us
with their experimental data. A.N.S. is grateful to M. da Mata and
is financially supported by FAPESP. M.W. acknowledges support by
the DAAD in the framework of the PROBRAL programme. M.W. and
A.N.S. thank L.G. Marcassa and V.S. Bagnato (USP S\~{a}o Carlos)
for their hospitality. The work of C-Y.L. was partially supported
by the National Science Council, ROC under the Grant
NSC-94-2112-M-259-008 and by FAPESP. C-Y.L. thanks the hospitality
of the members of Departamento de F\'{\i}sica Matem\'atica of USP,
where this work was performed.


\end{document}